\documentclass[aps,pra,twocolumn]{revtex4}%
\usepackage{amsmath}
\usepackage{amsfonts}
\usepackage{amssymb}
\usepackage{graphicx}%
\providecommand{\U}[1]{\protect\rule{.1in}{.1in}}
\begin{document}
\title{Quantum bit commitment under Gaussian constraints}
\author{Aikaterini Mandilara}
\affiliation{Quantum Information and Communication, \'{E}cole Polytechnique de Bruxelles,
CP~165/59, Universit\'{e} Libre de Bruxelles, 1050 Brussels, Belgium}
\author{Nicolas J. Cerf}
\affiliation{Quantum Information and Communication, \'{E}cole Polytechnique de Bruxelles,
CP~165/59, Universit\'{e} Libre de Bruxelles, 1050 Brussels, Belgium}
\affiliation{W.M. Keck Center for Extreme Quantum Information Theory, Massachusetts
Institute of Technology, 77 Massachusetts Avenue, Cambridge, MA 02139, USA}

\begin{abstract}
Quantum bit commitment has long been known to be impossible. Nevertheless,
just as in the classical case, imposing certain constraints on the power of
the parties may enable the construction of asymptotically secure protocols.
Here, we introduce a quantum bit commitment protocol and prove that it is
asymptotically secure if cheating is restricted to Gaussian operations. This
protocol exploits continuous-variable quantum optical carriers, for which such
a Gaussian constraint is experimentally relevant as the high optical
nonlinearity needed to effect deterministic non-Gaussian cheating is inaccessible.

\end{abstract}
\maketitle

\section{Introduction}

Quantum bit commitment (QBC) is probably one of most studied quantum
cryptographic primitive, just after quantum key distribution (see, e.g.,
\cite{DAriano}). It belongs to the class of mistrustful cryptography problems,
which involve two parties (Alice and Bob) who do not trust each other. More
specifically, bit commitment is a primitive in which Alice commits to a
certain bit while this bit should remain hidden to Bob until Alice later
reveals its value. In the first stage, called \textquotedblleft commit
phase\textquotedblright, Alice locks her bit in such a way that it is hidden,
and sends it to Bob. The protocol is said to be \textquotedblleft
concealing\textquotedblright\ if Bob cannot cheat by learning information
about this bit before the second stage. In this second stage, called
\textquotedblleft unveil phase\textquotedblright, Alice sends information to
Bob so that he can unlock the bit and find its value. The protocol is said to
be \textquotedblleft binding\textquotedblright if Alice cannot cheat by
changing the value of the bit once she has committed to it. A bit commitment
protocol is secure if it prevents Alice and Bob from cheating, that is, if it
is both binding and concealing.

The original proof of the impossibility of QBC due to Mayers consists of two
steps \cite{Mayers}. In the first, which is the most subtle one and will not
be discussed here, one shows that the security of any QBC reduces to the
security of a generic QBC scenario as described hereunder. The second step is
then to show that this generic QBC scenario, also known as a \textquotedblleft
purification\textquotedblright QBC protocol \cite{Spekkens}, is insecure
\cite{Lo}. In this scenario, Alice uses a bipartite Hilbert space
$H_{p}\otimes H_{t}$, which is the tensor product of the so-called
\textquotedblleft proof\textquotedblright and \textquotedblleft
token\textquotedblright spaces. Alice chooses to commit the bit $b$ ($0$ or
$1$) and prepares one of the two orthogonal states $\left\vert \chi
_{b}\right\rangle $ in the total Hilbert space by applying a unitary
transformation $U_{b}$ on a state $\left\vert \psi\right\rangle $, that is,
$\left\vert \chi_{b}\right\rangle =U_{b}\left\vert \psi\right\rangle $. In the
Schmidt representation, these states may be written as%

\begin{equation}
\left\vert \chi_{0}\right\rangle ={\displaystyle\sum\limits_{i}}%
a_{i}\left\vert p_{i}\right\rangle \left\vert t_{i}\right\rangle
,\qquad\left\vert \chi_{1}\right\rangle ={\displaystyle\sum\limits_{i}}%
a_{i}^{\prime}\left\vert p_{i}^{\prime}\right\rangle \left\vert t_{i}^{\prime
}\right\rangle \label{proof2}%
\end{equation}
In the commit phase, Alice transmits to Bob the token system lying in $H_{t}$,
which is in state $\rho_{b}=\mathrm{tr}_{p}|\chi_{b}\rangle\langle\chi_{b}|$.
In the unveil phase, Alice transmits to Bob the proof system lying in $H_{p}$,
so Bob can determine the value of the committed bit $b$ by projectively
measuring the state $\left\vert \chi_{b}\right\rangle $ using orthogonal
projectors. Now, the insecurity of this generic QBC protocol against cheating
can easily be proven. The requirement that Bob gains no information before the
unveil phase simply translates into $\rho_{1}=\rho_{0}$, or equivalently
$\left\vert t_{i}^{\prime}\right\rangle =\left\vert t_{i}\right\rangle $ (up
to a phase) and $a_{i}^{\prime}=a_{i}$, $\forall i$. Remarkably, if this
condition is fulfilled, Alice can perfectly cheat after the commit phase by
changing $\left\{  \left\vert p_{i}\right\rangle \right\}  \rightarrow\left\{
\left\vert p_{i}^{\prime}\right\rangle \right\}  $ with some appropriate
unitary transformations $U_{p}\otimes\openone$ on her proof system. This
implies that quantum bit commitment cannot be both perfectly concealing and
binding \cite{Mayers,Lo}.

This proof leaves open the possibility that if certain restrictions are
imposed on the operations available to the parties, a QBC may be constructed
that is secure or at least partially secure. There is some literature on this
topic for both classical and quantum bit commitment (see for instance
\cite{Terhal} and references therein, or \cite{Kent,Damgard,Cerf}), with
positive and negative results. What we shall examine here is a simpler and
less studied scenario, where restrictions are imposed on Alice's cheating
operations only. One can easily construct a secure QBC that falls into this
category: simply encode the committed bit into a subspace of the total Hilbert
space that remains invariant under Alice's permitted local transformations on
the proof system. To illustrate this idea, let us give a trivial example using
a system consisting of two spin-$1/2$ particles. Let us encode $0$ and $1$
into the eigenvalue of the total spin $\mathbf{S}^{2}=\left(  \mathbf{S}%
_{1}+\mathbf{S}_{2}\right)  ^{2}$ by choosing
\begin{align}
\left\vert \chi_{0}\right\rangle  &  =\left\vert s=0,m=0\right\rangle =\left(
\left\vert \uparrow\right\rangle \left\vert \downarrow\right\rangle
-\left\vert \downarrow\right\rangle \left\vert \uparrow\right\rangle \right)
/\sqrt{2}\nonumber\\
\left\vert \chi_{1}\right\rangle  &  =\left\vert s=1,m=0\right\rangle =\left(
\left\vert \uparrow\right\rangle \left\vert \downarrow\right\rangle
+\left\vert \downarrow\right\rangle \left\vert \uparrow\right\rangle \right)
/\sqrt{2}\label{ex2}%
\end{align}
where $s$ stands for the total spin quantum number and $m$ for the quantum
number associated with its projection onto the $z$ axis. It is obvious that
the condition $\rho_{1}=\rho_{0}$ is satisfied and that the protocol is not
secure if Alice has all local operations (i.e., the algebra $SU(2)\otimes1$)
at her disposal. However, let us suppose that her local operations are
restricted to a subgroup of $SU(2)\otimes1$ that commutes with $\mathbf{S}%
^{2}$. For the case of spins there is no such subgroup, but one can still
restrict Alice to use the trivial operation generated by $\mathbf{S}_{1}$,
that is, a rotation around the $(1,1,1)$ vector in the Bloch sphere
representation. Under this restriction, the protocol becomes secure since
cheating would require a rotation around the $z$ axis or $(0,0,1)$ vector,
i.e., an operation known as a \textquotedblleft phase gate\textquotedblright
in which $\left\vert \uparrow\right\rangle $ remains unchanged while
$\left\vert \downarrow\right\rangle $ gets a minus sign.

This example is rather unrealistic since there is no objective reason for
justifying this restriction on Alice's local operations while, in the total
Hilbert space $H_{p}\otimes H_{t}$, she can apply the global operations that
generate $\left\vert \chi_{b}\right\rangle $ as defined in Eq.~(\ref{ex2}). On
the contrary, in the QBC scheme that we introduce in this paper, it will
appear that a specific constraint on Alice's cheating operations can be
experimentally well motivated, giving rise to an asymptotically secure
protocol. We will devise a \textquotedblleft
continuous-variable\textquotedblright QBC protocol based on quantum states
lying in an infinite-dimensional Hilbert space, which can be realized as
states of the electromagnetic field (see, e.g., \cite{cvreview,book}). In this
quantum optical QBC protocol, we shall assume that Alice is restricted to
carry out Gaussian operations only, which is consistent with the current
experimental ability to engineer quantum states of light in a deterministic
way. Only a few very challenging experiments have been successful to prepare
and manipulate non-Gaussian states of traveling light (see, e.g.,
\cite{Polzik,Grangier1,Grangier2,Bellini,Furusawa1,Furusawa2}), and all of
these schemes are based on heralded photon subtraction \cite{subtraction} or
addition \cite{addition}, hence are probabilistic in nature. A deterministic
non-Gaussian operation would require high optical nonlinearities that are not
accessible in the laboratory today. Since probabilistic cheating does not
endeavor the security of QBC if the success probability is low (this even
holds true otherwise, though in the asymptotic protocol only), such a
restriction to Gaussian cheating operations is justified in the context of QBC.

Thus, although it is not impossible, in principle, to realize deterministic
non-Gaussian optical operations based on giant nonlinearities, there is a
natural boundary separating the Gaussian from non-Gaussian deterministic
operations, and it is relevant to investigate a QBC scenario where Alice is
not allowed to carry out non-Gaussian cheating operations. This scenario has
been introduced in \cite{Cerf}, where a strong \textquotedblleft no-go
theorem\textquotedblright\ was derived: secure quantum bit commitment is
forbidden in continuous-variable protocols where both players are restricted
to use Gaussian states and operations. In other words, if the protocol is
built on Gaussian states $\left\vert \chi_{b}\right\rangle $, it is sufficient
for the players to carry out Gaussian operations in order to cheat perfectly.
Therefore, it was concluded in \cite{Cerf} that a secure QBC protocol with
Gaussian constrained cheating, if it exists, should necessarily be built on
non-Gaussian states $\left\vert \chi_{b}\right\rangle $. In the present paper,
we prove that this holds by exhibiting an explicit secure non-Gaussian QBC
protocol. It should not be viewed as a directly usable QBC protocol since, as
we will see, it still requires the use of either a quantum memory or a very
long time delay in an optical interferometer. Instead, our goal is to point
towards a conceptual method to reach asymptotic security in
continuous-variable QBC under Gaussian constraints. A restricted
proof-of-principle demonstration of this protocol seems nevertheless feasible
within the currently available technologies.

In Section \ref{II}, we define our QBC protocol and analyze first how it works
when the two parties are honest. In Section \ref{III}, we go beyond the honest
scheme and investigate Alice's best possible cheating if restricted to
Gaussian operations. In Section \ref{IV}, we consider Bob's cheating, which
allows us to probe the trade-off between Alice and Bob's cheating. In Section
\ref{V}, we suggest an improvement to the scheme in order to make it
asymptotically secure, while we conclude in Section \ref{VI}.

\section{The honest scheme \label{II}}

Let us consider the following purification protocol \cite{Spekkens} in an
infinite-dimensional Hilbert space, which is in direct analogy with the
above-mentioned spin-$1/2$ example. The $0$ and $1$ values of the committed
bit $b$ are encoded into the orthogonal two-mode non-Gaussian states,
\begin{align}
\left\vert \chi_{0}\right\rangle  &  =\left(  \left\vert \alpha\right\rangle
\left\vert -\alpha\right\rangle -\left\vert -\alpha\right\rangle \left\vert
\alpha\right\rangle \right)  /\sqrt{2\left(  1-\mathrm{e}^{-4\left\vert
\alpha\right\vert ^{2}}\right)  }\nonumber\\
\left\vert \chi_{1}\right\rangle  &  =\left(  \left\vert \alpha\right\rangle
\left\vert -\alpha\right\rangle +\left\vert -\alpha\right\rangle \left\vert
\alpha\right\rangle \right)  /\sqrt{2\left(  1+\mathrm{e}^{-4\left\vert
\alpha\right\vert ^{2}}\right)  } \label{BB842}%
\end{align}
where $\left\vert \alpha\right\rangle =D\left(  \alpha\right)  \left\vert
0\right\rangle =\exp\left(  \alpha a^{\dag}-\alpha^{\ast}a\right)  \left\vert
0\right\rangle $ is a coherent state of complex amplitude $\alpha$. Moving
from orthogonal qubit states $\left\vert 0\right\rangle $ and $\left\vert
1\right\rangle $ in a two-dimensional Hilbert space to near-orthogonal
coherent states $\left\vert \alpha\right\rangle $ and $\left\vert
-\alpha\right\rangle $ ($\alpha\gg1$) in an infinite-dimensional Hilbert space
has already been put forward in the context of quantum computation
\cite{Jeong,Ralph}, and our treatment of QBC follows on this. Note that the
states of Eq.~(\ref{BB842}) correspond to entangled \textquotedblleft
Schr\"{o}dinger cat\textquotedblright states, whose experimental generation
has recently been demonstrated in \cite{Grangier3}.

%this on the security is in accordance with previous findings.

Let us suppose that $\alpha\gtrsim2$ (in practice, this is sufficient to be
very close to the asymptotic situation where $\left\vert \alpha\right\rangle $
and $\left\vert -\alpha\right\rangle $ are orthogonal). One of the two modes
(token system) of state $\left\vert \chi_{b}\right\rangle $ is sent to Bob in
the commit phase, while the second mode (proof system) is kept by Alice. At
this stage, Bob can almost not distinguish between $\left\vert \chi
_{0}\right\rangle $ and $\left\vert \chi_{1}\right\rangle $ whatever
measurement he uses since $\rho_{1}\simeq\rho_{0}$. On the other hand, it is
immediate to see how Bob can distinguish between these mutually orthogonal
states $\left\vert \chi_{b}\right\rangle $ in the total Hilbert space during
the unveil phase. Consider the two modes of $\left\vert \chi_{b}\right\rangle
$ as incident beams on the two ports of a balanced beam splitter, effecting
the unitary operation $B$. By adjusting the phases, the outgoing state
$\left\vert \chi_{b}^{\prime}\right\rangle =B\left\vert \chi_{b}\right\rangle
$ can be written as
\begin{align}
\left\vert \chi_{0}^{\prime}\right\rangle  &  =\left(  \left\vert
\alpha^{\prime}\right\rangle -\left\vert -\alpha^{\prime}\right\rangle
\right)  \left\vert 0\right\rangle /\sqrt{2\left(  1-\mathrm{e}^{-2\left\vert
\alpha^{\prime}\right\vert ^{2}}\right)  }\nonumber\\
\left\vert \chi_{1}^{\prime}\right\rangle  &  =\left(  \left\vert
\alpha^{\prime}\right\rangle +\left\vert -\alpha^{\prime}\right\rangle
\right)  \left\vert 0\right\rangle /\sqrt{2\left(  1+\mathrm{e}^{-2\left\vert
\alpha^{\prime}\right\vert ^{2}}\right)  } \label{cat2}%
\end{align}
which is the tensor product of a \textquotedblleft Schr\"{o}dinger
cat\textquotedblright state of amplitude $\alpha^{\prime}=\sqrt{2}\alpha$ and
the vacuum state. Note that the cat state is odd (even) for $b=0$ ($b=1$). The
states $\left\vert \chi_{b}^{\prime}\right\rangle $ are perfectly
distinguishable by applying a photon number parity measurement on the first
mode, corresponding to the observable $P=\left(  -1\right)  ^{a^{\dag}\!a}$,
that is,
\begin{equation}
\left\langle \chi_{0}^{\prime}\right\vert P\otimes1\left\vert \chi_{0}%
^{\prime}\right\rangle =-1,\qquad\left\langle \chi_{1}^{\prime}\right\vert
P\otimes1\left\vert \chi_{1}^{\prime}\right\rangle =1
\end{equation}
Note that this measurement may be realized by photon counting using a
number-resolving photodetector as has very recently become available (see,
e.g., \cite{Knill,Inoue})

\begin{figure}[h]
{\centering{\includegraphics*[ width=0.5\textwidth]{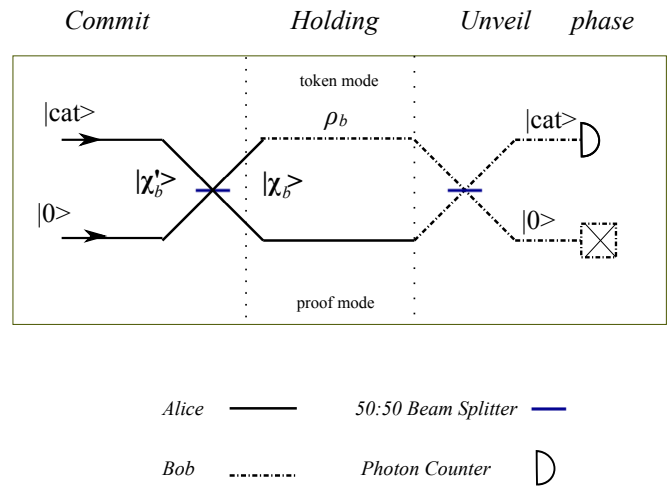}}} \vspace
{0.1cm}\caption{Honest protocol. The state $\left\vert cat\right\rangle $ can
either be an even or an odd Schr\"{o}dinger's cat state of amplitude
$\alpha^{\prime}$ depending on the bit $b$ to be committed, see the first mode
in Eq.~(\ref{cat2}). The token mode is transmitted in the commit phase, while
the proof mode is transmitted in the unveil phase. Bob combines the two modes
at a balanced beam splitter and measures the photon number parity in the first
mode.}%
\label{honestfig}%
\end{figure}

Now, we are ready to describe the honest QBC protocol as illustrated in Fig.
\ref{honestfig}. In the commit phase, Alice prepares one of the states
$\left\vert \chi_{b}^{\prime}\right\rangle $ as defined in Eq.~(\ref{cat2})
according to the value of the bit $b$ she wants to commit. Using a balanced
beam splitter, she converts $\left\vert \chi_{b}^{\prime}\right\rangle $ into
$\left\vert \chi_{b}\right\rangle $ as defined in Eq.~(\ref{BB842}), and then
sends the token mode (in state $\rho_{b}$) to Bob. In the unveil phase, she
sends the proof mode to Bob, which he combines with his token mode in a
balanced beam splitter to obtain the unentangled state $\left\vert \chi
_{b}^{\prime}\right\rangle $ as originally held by Alice. Finally, Bob
discards the mode corresponding to the vacuum state and performs a parity
measurement on the cat state in order to unveil the value of bit $b$. We
assume that the interferometric scheme is perfectly balanced and that the
holding phase (the period after the commit phase but before the unveil phase)
can be achieved by inserting equal time delays in the two branches of the
interferometer. Ideally, a quantum memory should of course be available to Bob
in order to achieve a longer-time holding phase.

\bigskip

\section{Alice's best cheating strategy \label{III}}

Let us first assume that the QBC protocol is concealing, that is, secure
against any measurement by Bob trying to cheat during the holding phase. In
other words, we assume that the coherent state amplitude $\alpha\gtrsim2$,
%entering Eqs.~(\ref{BB841})-(\ref{BB842}) are large enough to guarantee
so that $\rho_{1}\simeq\rho_{0}$. Under this assumption, which we will make
rigorous in Section~\ref{IV}, we can investigate the security of the protocol
against Alice's cheating strategies.

\subsection{Non-Gaussian cheating}

Obviously, if all local operations on the proof mode were available to Alice,
then she could convert the value of her committed bit at will during the
holding phase. However, this would require her ability to perform a notably
non-Gaussian local operation, where $\left\vert \alpha\right\rangle $ remains
unchanged while $\left\vert -\alpha\right\rangle $ gets a minus sign. Such an
operation, which can be viewed as the continuous-variable analogue to the
phase gate that we referred to in the case of spin-1/2 particles, corresponds
in the limit $\alpha\gg1$ to the non-Gaussian unitary
\begin{equation}
U_{NG}=D\left(  -\alpha\right)  \exp\left[  i \pi\left\vert 0 \right\rangle \!
\left\langle 0 \right\vert \right]  D\left(  \alpha\right)
\end{equation}
It cannot be implemented deterministically with accessible optical
nonlinearities, so for cheating one would have to turn to probabilistic
schemes based on heralded photon subtraction, whose probability of success is
very low \cite{Sasaki,Fiurasek}. Therefore, we may fairly impose such a
Gaussian restriction on Alice's cheating operation in the holding phase (or
assume that the probability of success of such a non-Gaussian local operation
is negligible).

One may rightly argue, of course, that within the current experimental
settings, the generation of Schr\"odinger cat states as needed in Alice's
preparation of $\left\vert \chi_{b}^{\prime}\right\rangle $ cannot be
deterministic either. However, during the commit phase, Alice can determine
whether $\left\vert \chi_{b}^{\prime}\right\rangle $ has been successfully
prepared or not, and, if not, she can repeat the operation again until it is
successful (or send the state to Bob but later notify him of the failed
trial). Thus, the probabilistic occurrence of a failure is not detrimental to
the commit phase, while it prevents an efficient non-Gaussian cheating. In
other words, the Gaussian restriction we impose on Alice's cheating operations
is justified within the present experimental limitations, while, at the same
time, Alice's preparation of the non-Gaussian state $\left\vert \chi
_{b}^{\prime}\right\rangle $ needed to overcome the no-go theorem \cite{Cerf}
can very well be done probabilistically.

\subsection{Gaussian cheating}

%Under the assumptions we have imposed,
The essential question to be answered now is to find the best cheating
strategy for Alice if only Gaussian operations are available to her. Here,
\textquotedblleft best\textquotedblright should be interpreted according to
the unveil procedure that has been defined for the honest protocol, that is,
when Bob uses a balanced beam splitter and performs a parity measurement on
the first mode. If, when Alice cheats, Bob reconstructs the state $|\chi
_{\#}^{\prime}\rangle$ at the output of his beam splitter while Alice had
initially committed $\left\vert \chi_{b}^{\prime}\right\rangle $, the best
cheating strategy is obviously the one where $\langle\chi_{\#}^{\prime
}|P\otimes1|\chi_{\#}^{\prime}\rangle$ reaches the closest value to
$\langle\chi_{\lnot b}^{\prime}|P\otimes1|\chi_{\lnot b}^{\prime}\rangle$,
where ${\lnot b}$ is the complement of the bit $b$. The probability of success
of the best Gaussian cheating strategy can be measured with Alice's
\textquotedblleft maximum control\textquotedblright$C_{\max}$, as defined in
\cite{Spekkens}.

Let us review the operations available to Alice. The most general Gaussian
unitary operation $U_{G}$ on a single mode is an exponential of a linear
combination of the elements of the two-photon algebra $h_{6}$, namely
$\left\{  \openone,a,a^{2},a^{\dag},a^{\dag^{2}},1/2+a^{\dag}a \right\}  $, so
it relies on $6$ real parameters. This general Gaussian transformation can
also be casted as a sequence of standard optical operations, for instance
\cite{Zhang}
\begin{equation}
U_{G}=D\left(  \beta\right)  U\left(  \varphi\right)  S\left(  r\right)
U\left(  \theta\right)  \label{Gaus}%
\end{equation}
where $S\left(  r\right)  =\exp\left[  \frac{r}{2} \left(  a^{2} - a^{\dag2}
\right)  \right]  $ is a squeezing of parameter r of the $x$ quadrature,
$U\left(  \theta\right)  =\exp\left(  \mathrm{i}\theta a^{\dag}a\right)  $ is
a phase rotation of angle $\theta$, and $D\left(  \beta\right)  =\exp\left(
\beta a^{\dag}-\beta^{\ast}a\right)  $ is a displacement of complex coherent
amplitude $\beta$. We ignore the global phase operation, which plays no role here.

The information about the committed bit $b$ is encoded into the parity of the
first mode of $\left\vert \chi_{b}^{\prime}\right\rangle $, or, adopting a
phase-space point of view, in the interference pattern of the Wigner function
($\hbar=1$),
\begin{equation}
W\left(  x,p\right)  =\frac{1}{2\pi}%
%TCIMACRO{\dint \nolimits_{-\infty}^{\infty}}%
%BeginExpansion
{\displaystyle\int\nolimits_{-\infty}^{\infty}}
%EndExpansion
\exp\left(  \mathrm{i}pq\right)  \left\langle x-\frac{q}{2}\right\vert
\rho\left\vert x+\frac{q}{2}\right\rangle \mathrm{d}q \label{Wigner}%
\end{equation}
Note that the quadrature variables $(x,p)$ in phase space are defined here
using the convention $a=(x+ip)/\sqrt{2}$. This phase-space interpretation,
which will be very useful in the following, originates from the relation
between the mean parity and the Wigner function at the origin in phase space,
namely
\begin{equation}
W\left(  0,0\right)  =\frac{1}{\pi}\langle P\rangle. \label{Parity}%
\end{equation}
The interference pattern of the Wigner function of the Schr\"{o}dinger cat
state $\left\vert \chi_{b}^{\prime}\right\rangle $ -- hence the parity
information -- is smoothed out during the commit phase since Alice looses a
handle on the token system, and it is revived in the unveil phase once the
token and proof systems can be measured jointly. One can visually understand
this smoothing out procedure by comparing the Wigner function of the first
mode in Eq.~(\ref{cat2}) during the commit phase (before the beam splitter)
with the Wigner function of the traced out mode in Eq.~(\ref{BB842}) during
the holding phase. What we need to analyze is the effect on the mean parity
$\langle P\rangle$ of the first mode of $\left\vert \chi_{b}^{\prime
}\right\rangle $ when Alice applies any Gaussian unitary $U_{G}$ on the proof
mode of $\left\vert \chi_{b}\right\rangle $. The most general operation is
actually a Gaussian CP map, but we will argue later on that the best cheating
is necessarily a Gaussian unitary.

In the simplest scenario, involving displacements only, Alice can for example
displace the proof system by $d$ along the positive $p$ quadrature direction.
In the unveiling phase, Bob will then get the initially committed cat state
displaced by $-d/\sqrt{2}$ along the $p$ quadrature, where the factor
$\sqrt{2}$ is due to the second beam splitter. In other words, Alice can alter
the parity of the unveiled state by freely displacing the origin of
phase-space to another point of the interference pattern where the Wigner
function has another value, even possibly the opposite sign. We will now prove
that this simplest scenario actually provides the best Gaussian cheating
strategy for Alice, so that no squeezing or phase-rotation is helpful.

%We  explain this statement here using qualitative
%arguments since the accurate proof is straighforward and tedious.

%leaving displacement aside, any Gaussian unitary corresponds
%to a symplectic transformation in phase-space, which is actually a special case of an
%affine affine transformation.

The key observation is that the most general Gaussian unitary of
Eq.~(\ref{Gaus}) corresponds to a special case of an affine transformation in
phase-space \cite{Schleich}, namely a linear symplectic transformation
followed by a translation. Intuitively, this means that the Wigner function of
the initial state may be displaced, squeezed, or rotated, but its maximum and
minimum values $W_{\max,\min}$ remain invariant under these operations.
Alternatively, using Eq.~(\ref{Parity}), this means that the maximum and
minimum values of the mean parity $\langle\hat{P}\rangle_{\max,\min}$ that can
be reached under Gaussian unitaries are invariant for a given input state.
They can be reached simply by translating the Wigner function in such a way
that the origin is moved towards the highest peak or the deepest dip in phase
space, respectively, with no squeezing or rotation needed.

Remember that, when cheating, Alice can only apply her Gaussian operation on
the proof mode, not on the token mode. However, since Bob only checks the
first outgoing mode of his beam splitter (the one containing the cat state
whose parity encodes $b$), an arbitrary displacement on this mode can be
achieved via a displacement of the proof mode only, so Alice can indeed freely
translate the Wigner function of the cat state. Now, leaving displacements
asides, if Alice's cheating operation involves a rotation or squeezing
operation, the outgoing modes of Bob's beam splitter become inevitably
entangled, so the unveiled state becomes mixed. Since mixing can only wash out
the interference pattern, the maximum parity $\langle P\rangle_{\max}$ can
only decrease while $\langle P\rangle_{\min}$ can only increase. Thus,
rotation and squeezing can only make cheating worse, and are useless to Alice.
The same reasoning also implies that a Gaussian CP map cannot do better than a
Gaussian unitary since it eventually implies tracing over some ancillary
system after applying a Gaussian unitary onto the joint system, hence smearing
out the Wigner function.

%In our case though, a Gaussian operation
%on the proof mode does not necessary implies a Gaussian operation on the
%exit-Bob modes, apart from the special case where the Gaussian operation is
%just a displacement.

This confirms that Alice's best Gaussian cheating strategy for reaching the
target bit value
%$\left\langle \hat{P}\otimes1\right\rangle _{\min(\max)}$
%$\langle \chi_{\neg b}^{\prime}\vert \hat{P}\otimes1\vert \chi_{\neg b}^{\prime}\rangle $
$\neg b = 0$ $(1)$ is by displacing her proof system so that Bob obtains the
originally committed cat state $\left\vert \chi_{b}^{\prime}\right\rangle $
displaced in such a way that the minimum (maximum) value of its Wigner
function $W_{\min}$ ($W_{\max}$) is now located at the origin. Perfect
cheating will be achieved if $W_{\min}=-1$ ($W_{\max}$=1).

\subsection{Alice's maximum control $C_{\max}$}

\begin{figure}[h]
{\centering{\includegraphics*[ width=0.5\textwidth]{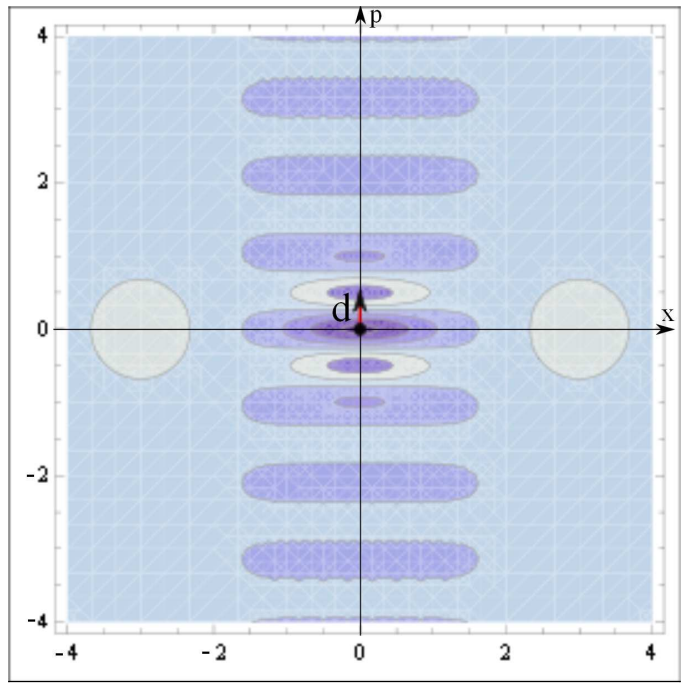}}} \vspace
{0.1cm}\caption{Contour plot of the Wigner function of an odd Schr\"{o}dinger
cat state of amplitude $\alpha^{\prime}=3/\sqrt{2}$. Bob obtains this state in
the unveil phase if Alice has committed the bit $0$ and has not cheated. The
best cheating strategy for Alice during the holding phase is to displace her
proof system by $\sqrt{2}d$ along the $p$ quadrature, so that the unveiled
state is displaced by $-d$ (or the origin in phase space is shifted upwards by
$d$, as illustrated by the arrow). Then, the origin of the unveiled state is
located on the global maximum of the \ original Wigner function. For an even
Schr\"{o}dinger cat, the situation is exactly analogous.}%
\label{Fig2}%
\end{figure}

To illustrate this optimal Gaussian cheating, suppose that Alice has initially
committed the bit $b=0$ (odd Schr\"{o}dinger cat with $\langle\chi_{0}%
^{\prime}|P\otimes1|\chi_{0}^{\prime}\rangle=-1$)
%($\left\langle \hat{P}\otimes1\right\rangle =-1$)
and attempts to cheat during the holding phase so that Bob would measure a bit
$\lnot b=1$ (even Schr\"{o}dinger cat with $\langle\chi_{\#}^{\prime}%
|P\otimes1|\chi_{\#}^{\prime}\rangle=1$)
%($\left\langle \hat{P}\otimes 1\right\rangle =1$),
in the unveil phase. The optimal cheating strategy is easy to understand in
Fig.~\ref{Fig2}, where we have plotted the Wigner function of the initially
committed cat state. Alice needs to displace her proof system by $\sqrt{2}\,d$
along the $p$ quadrature, so that $|\chi_{\#}^{\prime}\rangle$ becomes the
original odd cat state displaced by $-d$ along the $p$ quadrature
(equivalently, the origin of phase space is shifted upwards by $d$ as
illustrated by an arrow in Fig.~\ref{Fig2}). The parameter $d$ is just the
distance (along the $p$ quadrature direction) from the origin to the first
maximum of the interference pattern, which is also the global maximum of the
Wigner function.

%In terms of equations, the best cheating strategy is
%\begin{equation}
%D\left(  id\left(  \alpha^{\prime}\right)  \right)  \rho_{b}D^{\dag}\left(
%id\left(  \alpha^{\prime}\right)  \right)  \label{BestCh}%
%\end{equation}
%where $\alpha^{\prime}$ is the amplitude of the commited cat state,
%Eqs.(\ref{cat1})-(\ref{cat2}).

\begin{figure}[h]
{\centering{\includegraphics*[ width=0.5\textwidth]{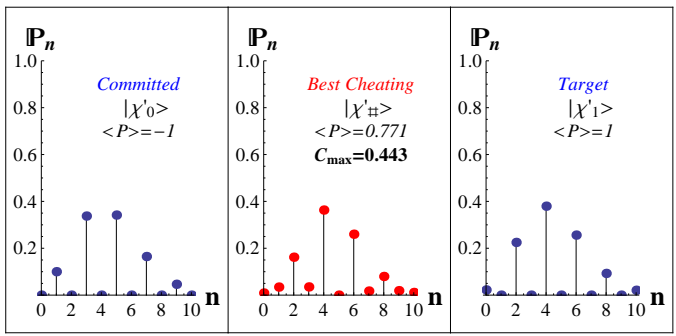}}} \vspace
{0.1cm}\caption{Photon number distribution for (a) the committed cat state
($\alpha^{\prime}=3/\sqrt{2}$) corresponding to bit $0$; (b) the state
achieved by Alice's best Gaussian cheating strategy; and (c) the target state
corresponding to bit $1$. The symbol $\left\langle P\right\rangle $ denotes
the mean photon number parity, while $C_{\max}$ denotes Alice's maximum
control.}%
\label{Fig3}%
\end{figure}

For a cat state of amplitude $\alpha^{\prime}$ there is no analytical
expression for $d$ as a function of $\alpha^{\prime}$, and one has to
numerically solve the equation
\begin{equation}
p\cos\left(  2\sqrt{2}p\alpha^{\prime}\right)  +\sqrt{2}\alpha\sin\left(
2\sqrt{2}p\alpha^{\prime}\right)  =pe^{-2\alpha^{\prime2}} \label{d}%
\end{equation}
for $p$. The best cheating is then a displacement by $d$, which corresponds to
the smallest positive and non-zero solution of Eq.~(\ref{d}). Let us analyze
precisely the effect of such a cheating in the specific example of
Fig.~\ref{Fig2}, that is, when Alice commits an odd cat state of amplitude
$\alpha^{\prime}=3/\sqrt{2}$. By solving Eq.~(\ref{d}), we get that the best
Gaussian cheating requires a displacement of $d=0.496$. The corresponding
photon number distributions with and without cheating are schematically
presented in Fig.~\ref{Fig3}, where we observe that the distribution with
cheating qualitatively resembles the target distribution.

The probability of success of this best Gaussian cheating strategy can be
measured with Alice's \textquotedblleft maximum control\textquotedblright%
$C_{\max}$ as defined in \cite{Spekkens}, that is, the largest difference
between Alice's probability of unveiling whatever bit she wants when she is
cheating and when she is honest. Assuming that the bit she wishes to unveil is
equiprobable, this can be expressed as one half of her cheating probability.
Using the relations $\left\langle P\right\rangle =\left\langle P_{+}%
\right\rangle -\left\langle P_{-}\right\rangle $ and $\left\langle
P_{+}\right\rangle +\left\langle P_{-}\right\rangle =1$, where $\langle
P_{+(-)}\rangle$ stands for the probability to measure an even (odd) number of
photons at the displaced origin, we can deduce $\langle P_{+(-)}\rangle$ from
the mean parity at the displaced origin $\left\langle P\right\rangle
=\pi\,W\left(  0,d\right)  $. Since Alice had committed a bit $b=0$, here
$\langle P_{+}\rangle$ is the probability that she successfully cheats and
unveils a bit $b=1$. Thus, in the present case, we get
\begin{equation}
C_{\max}\equiv\frac{1}{2}\left\langle P_{+}\right\rangle =\frac{1}{4}\left(
\left\langle P\right\rangle +1\right)  =0.443
\end{equation}

\begin{figure}[h]
{\centering{\includegraphics*[ width=0.37\textwidth]{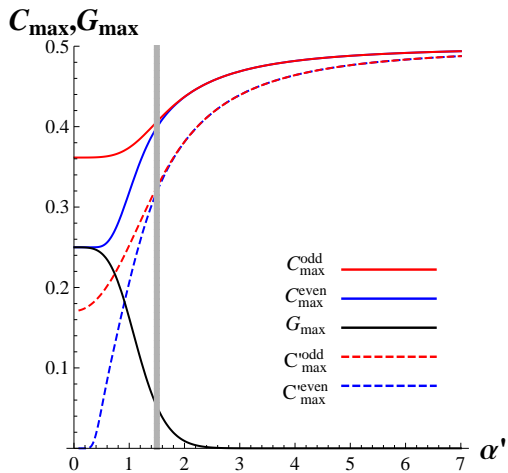}}}
\vspace{0.1cm}\caption{ Alice's maximum control $C_{\max}$ (one half of the
success probability of her optimal Gaussian cheating) as a function of the
coherent amplitude $\alpha^{\prime}$. For small values of $\alpha^{\prime}$
($\alpha^{\prime}\lesssim3/2$), on the left of the vertical grey bar, the
committed even and odd cat states behave differently, that is, $C_{\max
}^{\mathrm{odd}} \neq C_{\max}^{\mathrm{even}}$. We also plot $C_{\max
}^{\prime}$, Alice's maximum control when the vacuum mode is also monitored by
Bob in the unveil phase, for both an initially committed even and odd cat
state. Bob's maximum information gain $G_{\max}$ (one half of the probability
of successfully determining the committed bit) is also plotted as a function
of $\alpha^{\prime}$. }%
\label{Fig4}%
\end{figure}

The success probability of Alice's optimal Gaussian cheating $C_{\max}$
increases with the amplitude $\alpha^{\prime}$ since the contrast in the
interference pattern of the Wigner function of the cat state becomes stronger.
In Fig.~\ref{Fig4}, we illustrate this dependence as derived numerically. Note
that for small values of $\alpha^{\prime}$, there is a different behavior for
odd and even cats related to the fact that their mean photon number
significantly differs (the even cat tends to the vacuum state $\left\vert
0\right\rangle $ as $\alpha^{\prime}\rightarrow0$, while the odd cat tends to
the first number state $\left\vert 1\right\rangle $).
%In this parametric area ($\alpha^{\prime}\lesssim3/2$) the two possibly committed cat
%states differ in their average number of photons and therefore our protocol
%should no  be consider there, even if for completeness we include this area in
%the graphs.
For large values of $\alpha^{\prime}$ ($\gtrsim3/2$), it can be analytically
shown that the dependence simply scales as $C_{\max} \simeq\exp( -\pi
^{2}/8\alpha^{\prime2})/2 $, that is, $C_{\max}$ tends to 1/2 with a
difference following a polynomial dependence in $1/\alpha^{\prime}$. This
feature will be crucial in Sec.~\ref{V}, where we consider the asymptotic
security of the protocol. It is simply obtained by using the approximation
$d\approx\pi/(2\sqrt{2}\alpha^{\prime})$, resulting from the fact that in this
limit the global maximum (minimum) approximately coincides with the maxima
(minima) of the oscillating interference term $\cos\left(  2\sqrt{2}%
p\alpha^{\prime}\right)  $.

Interestingly, Alice's maximum control $C_{\max}$ can be further reduced if,
during the unveil stage, Bob also verifies that the second outgoing mode of
his beam splitter is in the vacuum state $\left\vert 0\right\rangle $ as it
should be in the absence of cheating. If Alice applies the above optimal
Gaussian cheating during the holding phase, the second outgoing mode
experiences the same displacement as the cat state, so that the probability
that no photon is detected is $P_{\mathrm{NoP}}=\exp\left(  -d^{2}/2\right)
$. This was irrelevant in the above calculation of $C_{\max}$ as Bob
disregarded the second mode in the unveil phase. However, one can make use of
this fact and measure the second mode with photon counting in order to further
improve the security of the protocol since the probability of successful
cheating is then reduced to $C_{\max}^{\prime}=C_{\max}\times P_{\mathrm{NoP}%
}$. We present the modified curve $C_{\max}^{\prime}$ in Fig.~\ref{Fig4} with
dashed lines. It must be stressed, however, that the optimal cheating strategy
we derived in the original protocol (without monitoring the second mode) does
not necessarily remain optimum for this modified protocol. Finding the optimal
$C_{\max}^{\prime}$ is an open problem.

It is important to mention here that throughout our analysis, we have only
considered the case where Alice applies her cheating operations during the
holding phase. Alice could as well cheat in the commit phase already by
commiting a state that is different from the state $\left\vert \chi
_{b}^{\prime}\right\rangle $ of Eq.~(\ref{cat2}) and try to change it during
the hold phase. An example of such a successful cheating strategy would be to
commit a state of amplitude $\alpha$ higher than the one agreed upon and
displace her proof mode at will during the commit phase. In this way, and if
$\alpha\gg1$, she could obviously achieve $C_{\max}$ asymptotically close to
$1/2$. However, the detection of such a cheating during the commit phase could
easily be detected in the asymptotic protocol that we suggest in Sec.~\ref{V}.

\section{Bob's best cheating strategy \label{IV}}

\begin{figure}[h]
{\centering{\includegraphics*[ width=0.4\textwidth]{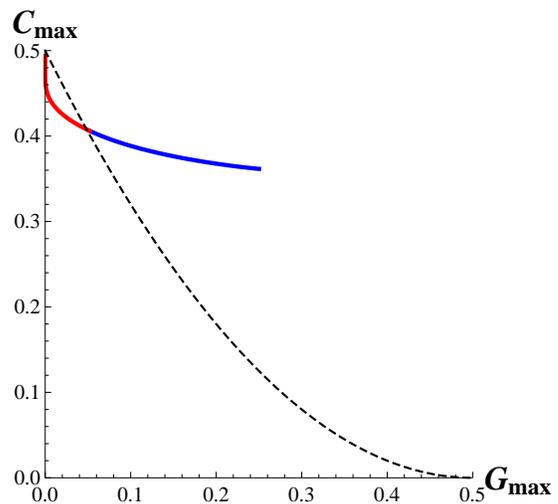}}} \vspace
{0.1cm}\caption{\textit{Solid red (blue) line}: Alice's maximum control
$C_{\max-Odd}$ versus Bob's maximum information gain $G_{\max}$ for
$\alpha^{\prime}\geq3/2$ ($\alpha^{\prime}<3/2$).\textit{ Dotted black line}:
Lower bound on the $C_{\max}$ versus $G_{\max}$ trade-off for QBC protocols as
derived in \cite{Spekkens}.}%
\label{Fig5}%
\end{figure}

In the previous Section, we have assumed that the amplitude $\alpha$ was large
enough to guarantee that the protocol was perfectly concealing. We will now
make this statement more accurate, and determine the relation between $\alpha$
and Bob's maximum information gain $G_{\max}$ (as defined in \cite{Spekkens})
during the holding phase while assuming that Alice is honest. The most
appropriate measure to quantify $G_{\max}$ uses the trace distance
\cite{Fucks}
\begin{equation}
D\left(  \rho_{0},\rho_{1}\right)  =\frac{1}{2}\mathrm{Tr}\left\vert \rho
_{0}-\rho_{1}\right\vert \label{Kolmo}%
\end{equation}
which corresponds to the probability of successfully distinguishing the two
quantum states with the best POVM measurement, so-called Helstrom measurement
\cite{Helstrom}. If $\rho_{0}$ and $\rho_{1}$ correspond to the single-mode
reduced states of $\left\vert \chi_{0}\right\rangle $ and $\left\vert \chi
_{1}\right\rangle $ from Eq.~(\ref{BB842}), it is easy to show that
\begin{align}
\rho_{0}-\rho_{1} &  =\frac{\mathrm{e}^{4\left\vert \alpha\right\vert ^{2}}%
}{\left(  \mathrm{e}^{8\left\vert \alpha\right\vert ^{2}}-1\right)  }\left(
\left(  \left\vert \alpha\right\rangle \left\langle \alpha\right\vert
+\left\vert -\alpha\right\rangle \left\langle -\alpha\right\vert \right)
\right.  \nonumber\\
&  \left.  -\mathrm{e}^{2\left\vert \alpha\right\vert ^{2}}\left(  \left\vert
\alpha\right\rangle \left\langle -\alpha\right\vert +\left\vert -\alpha
\right\rangle \left\langle \alpha\right\vert \right)  \right)  .\label{rho01}%
\end{align}
It remains to find the eigenvalues of the Hermitian matrix $H=\rho_{0}%
-\rho_{1}$, which is not a difficult task once we observe that it can be
rewritten in terms of cat states,
\begin{equation}
H=\lambda_{+}\left\vert +\right\rangle \left\langle +\right\vert +\lambda
_{-}\left\vert -\right\rangle \left\langle -\right\vert
\end{equation}
with $\left\vert \pm\right\rangle =\left(  \left\vert \alpha\right\rangle
\pm\left\vert -\alpha\right\rangle \right)  /\sqrt{2\left(  1\pm
\mathrm{e}^{-2\left\vert \alpha\right\vert ^{2}}\right)  }$ and $\lambda_{\pm
}=$ $\pm\mathrm{e}^{2\left\vert \alpha\right\vert ^{2}}/\left(  1+\mathrm{e}%
^{4\left\vert \alpha\right\vert ^{2}}\right)  $. It then follows that
\begin{equation}
\mathrm{Tr}|H|=2\mathrm{e}^{2\left\vert \alpha\right\vert ^{2}}/\left(
1+\mathrm{e}^{4\left\vert \alpha\right\vert ^{2}}\right)
\end{equation}
%\left(  \left\vert \alpha\right\rangle \left\langle
%-\alpha\right\vert +\left\vert -\alpha\right\rangle \left\langle
%\alpha\right\vert \right)
and%
\begin{equation}
G_{\max}\equiv\frac{1}{2}D\left(  \rho_{0},\rho_{1}\right)  =\frac
{e^{-2\left\vert \alpha\right\vert ^{2}}}{2\left(  1+\mathrm{e}^{-4\left\vert
\alpha\right\vert ^{2}}\right)  }.\label{Di}%
\end{equation}
This implies that Bob's capability to optimally distinguish between the two
states decays exponentially with $\alpha$, analogously to the behavior of the
overlap $\left\langle \alpha\right.  \left\vert -\alpha\right\rangle $. In
Figure~\ref{Fig4}, we have also plotted $G_{\max}$ as a function of the
amplitude of the committed cat state $\alpha^{\prime}=\sqrt{2}\alpha$, which
illustrates this exponential decay. We observe a trade-off between Alice's
cheating and Bob's cheating: the more control Alice has on the committed state
(large $C_{\max}$), the less information Bob is able to gain (small $G_{\max}%
$). This trade-off is exhibited in Fig. \ref{Fig5}, where we plot $C_{\max}$
versus $G_{\max}$ for an initially commiteed odd cat state (bit 0). For
comparison, we also plot the (not necessary reachable) lower bound on this
trade-off for QBC protocols as derived in \cite{Spekkens} (we refer the reader
to Ref.~\cite{Kerenidis} for some recently obtained results on the exact
bounds). It appears that, while our protocol as such cannot be both perfectly
concealing and binding (there is no value of $\alpha$ such that $G_{\max}$ and
$C_{\max}$ tend to zero simultaneously), it enters (for $\alpha^{\prime
}\geqslant3/2$) in the area that is not accessible to QBC protocols without restrictions.

\section{Asymptotically secure protocol \label{V}}

The fact that Bob's maximum information gain $G_{\max}$ is
\textit{exponentially} decreasing with $\alpha^{\prime}$ while the success
probability of Alice's best Gaussian cheating $C_{\max}$
%(or $C_{\max}^{\prime}$)
is only \textit{polynomially} increasing with $\alpha^{\prime}$ can be
exploited to improve the security of our QBC protocol in a similar manner as
in the original QBC protocol of Ref.~\cite{BB84}.

This is achieved by modifying the setting of Fig.~\ref{honestfig} and use a
sequence (a tensor product) of $N$ identical states $\left\vert \chi
_{b}^{\prime}\right\rangle $ instead of a single one for the encoding. In this
modified scheme, we may assume that Alice's best Gaussian cheating strategy
factorizes. A collective Gaussian attack on $N$ states cannot increase the
maximum value of the total Wigner function of the $N$ states, and therefore it
cannot give a better cheating on average. With this argument, we can estimate
that her maximum success probability is simply $C_{\max}^{(N)}=2^{N-1}\left(
C_{\max}\right)  ^{N}$ since the cheating remains undetected only if all $N$
states are successfully controlled by Alice.
%with $C_{\max}=C_{\max}^{(N=1)}$,
Hence $C_{\max}^{(N)}$ decreases \textit{exponentially} with $N$. In contrast,
assuming that entangled measurements are of no use, Bob's maximum information
gain becomes $G_{\max}^{(N)}=\left(  1-(1-2G_{\max})^{N}\right)  /2$ since
$(1-2G_{\max})$ is the probability of not distinguishing the states. Hence,
$G_{\max}^{(N)}$ increases linearly (at most polynomially \cite{Fucks}) with
the number of states $N$ provided that $G_{\max}$ is small ($\alpha^{\prime}$
is large).

Then, by choosing a large amplitude $\alpha^{\prime}$ so that $G_{\max}$ is
exponentially small, $C_{\max}^{(N)}$ can be made exponentially small as well
by choosing a large enough $N$ (not too large to keep $G_{\max}^{(N)}$ small,
which is possible given the linear scaling). In this way, we can construct a
QBC protocol that is \textit{asymptotically secure} in the sense that
$G_{\max}^{(N)}\rightarrow0$ and $C_{\max}^{(N)}\rightarrow0$.

\begin{figure}[h]
{\centering{\includegraphics*[ width=0.4\textwidth]{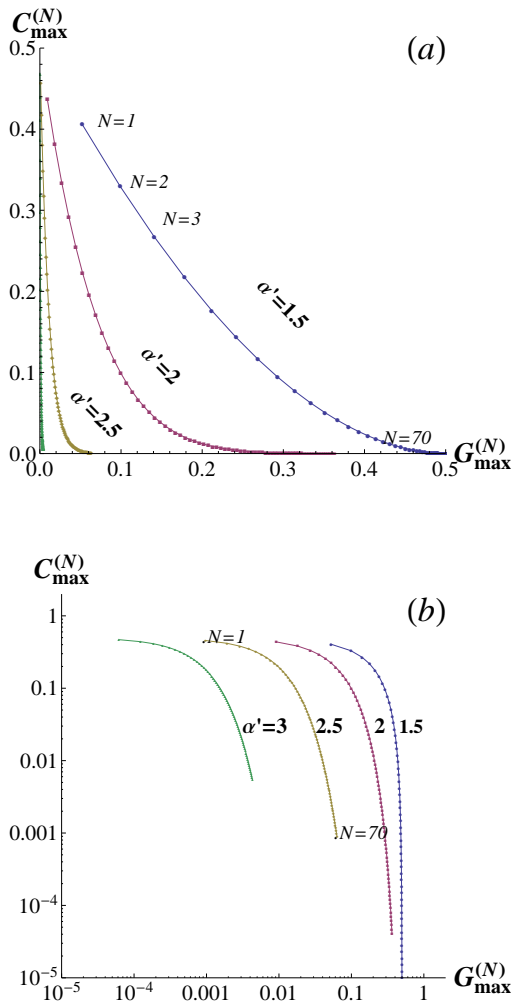}}} \vspace
{0.1cm}\caption{\textit{(a)} Alice's maximum control $C_{\max}^{(N)}$ versus
Bob's maximum information gain $G_{\max}^{(N)}$ for different amplitude
$\alpha^{\prime}$. (b) The same plot in logarithmic axes.}%
\label{Fig6}%
\end{figure}

As a matter of concreteness, we plot in Fig.~\ref{Fig6} the value of $C_{\max
}^{(N)}$ versus $G_{\max}^{(N)}$ for different values of the coherent
amplitude $\alpha^{\prime}$. For a given $\alpha^{\prime}$, the point moves to
the right for increasing $N$, and we tend to a protocol where Alice cannot
cheat anymore while Bob is able to cheat perfectly. If the value of
$\alpha^{\prime}$ is increased, the starting point for $N=1$ corresponds to a
better control for Alice and a lower information gain for Bob. Then, if
$\alpha^{\prime}$ is taken large enough, we can reach an interesting region
where both $C_{\max}^{(N)}$ and $G_{\max}^{(N)}$ are small by choosing an
appropriate large value of $N$.

In practice, achieving really small values of $G_{\max}$ and $C_{\max}$ is
probably not possible within the current available technology. For instance, a
security of the order of $10^{-5}$ would requires $\alpha^{\prime}\approx4$
and $N\approx300$. More realistically, a value of $G_{\max}$ and $C_{\max}$ of
the order of $10^{-1}$ would only require $\alpha^{\prime}\approx2$ and
$N\approx10$, which may be feasible if the $N$ committed cats states are sent
iteratively. For $N=1$, as can be seen in Fig. \ref{Fig5}, we are far from the
secure region as $C_{\max}$ remains too large. It may be interesting, however,
to demonstrate this protocol for $N=1$ and $\alpha^{\prime}\geqslant3/2$ as it
then beats any possible QBC protocol with no restriction \cite{Spekkens}, as
already mentioned.

The asymptotic protocol can also efficiently protect against cheating
strategies of Alice during the commit phase if the parity measurement at the
unveil phase is replaced by photon-number counting. In this case, by measuring
$N\gg1$ states, Bob obtains the photon-number distribution of the commited
state and thus may easily conclude if Alice has initially commited another
state than than the states $\left\vert \chi_{b}^{\prime}\right\rangle $ of
Eq.~(\ref{cat2}). For instance, in the case where Alice decides to commit \ a
state of amplitude $\alpha$ higher than the one agreed upon, the photon-number
distribution obtained by Bob will have a mean that is higher than expected.

Note finally that if the vacuum mode is monitored in the unveil phase and if
for this modified protocol the Gaussian cheating strategy we have examined is
proven to be the optimum, then $C_{\max}$ may be further reduced by a
significant factor. The maximum information gain $G_{\max}$ is also expected
to be, in practice, less than the values we have calculated since Helstrom
measurements for continuous variables require the use of non-Gaussian
resources. Finding an operational measurement scheme realizing the POVM
described in Section \ref{IV} or finding a (more convenient) tomographic
procedure for $N\gg1$ that achieves the maximum information gain is another
open question.

\section{Conclusions \label{VI}}

We have investigated continuous-variable QBC protocols with Gaussian
constraints. It had been proven in a recent work \cite{Cerf} that restricting
both parties to Gaussian states and operations cannot lead to a secure QBC
protocol. Here, we have gone one step further and have introduced a QBC
protocol that is based on non-Gaussian (Schr\"{o}dinger cat) states of light,
thereby circumventing such a Gaussian no-go theorem, but that still imposes a
Gaussian restriction on Alice's cheating operations. This continuous-variable
QBC protocol is shown to be asymptotically secure in the sense that Alice's
control and Bob's information gain can be both made arbitrarily small. Even
though the Gaussian restriction we put on Alice is not of a fundamental
nature, the non-Gaussian deterministic operations as needed by Alice in order
to cheat would require high optical non-linearities that are inaccessible
today in the laboratory. In contrast, the probabilistic procedures that can
effect non-Gaussian operations based on post-selection, as already
demonstrated in the laboratory, can be used by Alice in order to prepare the
cat states that are necessary to initiate the protocol.

In conclusion, we envision that a restricted proof-of-principle demonstration
of this continuous-variable QBC protocol may become realizable within the
near-future state of technology given the recent experimental progress on
non-Gaussian state of light generation \cite{Grangier3,Knill,Inoue}. An
interesting extension of this work would be to devise more practical
continuous-variable QBC protocols going beyond the purification protocol
investigated here, but instead following the lines of the original QBC
protocol of Ref.~\cite{BB84} for discrete variables, for which no entanglement
or quantum memory is required.

%the restrictions that we impose on Alice's local operations are well justified
%by the present experimental limitations, but, at the same time, the preparation
%of non-Gaussian cat states needed to overcome the no-go theorem can be done
%in a probabilistic scheme, so

%In both cases, the succesful outcome  is probabilistic and our proposed protocol seem to remains asymptotically secure
%even if the Gaussian restriction is relaxed.

We are grateful to J.~Fiurasek, X.~Lacour, and L.~Magnin for many useful
discussions. AM gratefully acknowledges financial support from the Belgian
National Fund for Scientific Research (F.R.S.-FNRS). This work was carried out
with the financial support of the European Commission via the project COMPAS,
the support of the National Fund for Scientific Research (F.R.S.-FNRS) via the
EraNet project HIPERCOM, and the support of the Brussels-Capital Region via
the project CRYPTASC.

\end{document}